\documentclass[10pt,twocolumn,english,aps,prl]{revtex4-1}
\usepackage{lmodern}
\usepackage{soul}
\setstcolor{blue}
\usepackage[T1]{fontenc}
\usepackage{color}
\usepackage{float}
\usepackage{amsmath}
\usepackage{amssymb}
\usepackage{graphicx}
\usepackage{esint}
\usepackage{bm}

\makeatletter

\renewcommand{\vec}{\mathbf}
\renewcommand{\vr}{\vec{r}}

\newcommand{\vb}{\vec{b}}

\newcommand{\vq}{\vec{q}}

\newcommand{\vx}{\vec{x}}

\newcommand{\vsigma}{\mbox{\boldmath $\sigma$}}

\newcommand{\vk}{\vec{k}}

\newcommand{\KK}{K}
\newcommand{\KKc}{K_{\rm c}}

\@ifundefined{textcolor}{}
{%
 \definecolor{BLACK}{gray}{0}
 \definecolor{WHITE}{gray}{1}
 \definecolor{RED}{rgb}{1,0,0}
 \definecolor{GREEN}{rgb}{0,1,0}
 \definecolor{BLUE}{rgb}{0,0,1}
 \definecolor{CYAN}{cmyk}{1,0,0,0}
 \definecolor{MAGENTA}{cmyk}{0,1,0,0}
 \definecolor{YELLOW}{cmyk}{0,0,1,0}
}

\makeatother

\usepackage{babel}
\begin{document}

\title{Quantum Transport of Disordered Weyl Semimetals at the Nodal Point}
\begin{abstract}
Weyl semimetals are paradigmatic topological gapless phases in three dimensions. We here address the effect of disorder on charge transport in Weyl semimetals. For a single Weyl node with energy at the degeneracy point and without interactions, theory predicts the existence of a critical disorder strength beyond which the density of states takes on a nonzero value. Predictions for the conductivity are divergent, however. In this work, we present a numerical study of transport properties for a disordered Weyl cone at zero energy. For weak disorder our results are consistent with a renormalization group flow towards an attractive pseudoballistic fixed point with zero conductivity and a scale-independent conductance; for stronger disorder diffusive behavior is reached. We identify the Fano factor as a signature that discriminates between these two regimes.      
\end{abstract}

\author{Bj\"orn Sbierski, Gregor Pohl, Emil J. Bergholtz, Piet W. Brouwer}

\affiliation{Dahlem Center for Complex Quantum Systems and Institut f\"ur Theoretische
Physik, Freie Universit\"at Berlin, D-14195, Berlin, Germany}

\date{July 11th, 2014}

\maketitle
\emph{Introduction.---}
Topological considerations not only can be used to describe and classify 
band insulators and superconductors \cite{Hasan2010,Bernevig2013}, 
they were also found to apply to gapless phases of matter 
\cite{Wen2002,Horava2005,Sato2006,Beri2010,Volovik2011,Bernard2012,Zhao2013,Matsuura2013}.
Perhaps the best known example of
a topologically nontrivial gapless band structure is that of
graphene \cite{Wallace1947}, which has four topologically
protected band touchings. The paradigmatic example of a 
topological gapless phase in three dimensions is the Weyl semimetal
\cite{Murakami2007, Wan2011, *[{For a review see }]  [{}] Hosur2013}, 
which features pairs of topologically protected gap closing points
in its Brillouin zone. The dispersion in the vicinity of a single
isotropic nodal point can be described by the effective Hamiltonian
\begin{equation}
  H_{0}\left(\mathbf{k}\right)= \pm \hbar 
  v\boldsymbol{\sigma}\cdot\mathbf{k}\label{eq:H0},
\end{equation}
where $v$ is the Fermi velocity, $\boldsymbol{\sigma}$ the vector
of Pauli matrices, $\pm$ denotes the chirality,
and $\mathbf{k}$ measures the Bloch wavevector
relative to the momentum in the Brillouin zone at which the gap
closing appears. 

Weyl semimetals have attracted considerable attention
due to the prediction of protected surface states with a Fermi arc
\cite{Wan2011} and the chiral anomaly in electromagnetic response
\cite{Goswami2013,*Vazifeh2013}. An ideal Weyl semimetal with
Fermi energy at the Weyl point $\varepsilon = 0$ has a vanishing
conductivity $\sigma$, but a finite conductance \cite{Baireuther2014}, making it neither conducting nor insulating.
The excitement is further fueled by the existence of concrete
theoretical proposals for material candidates for Weyl semimetals, both in the solid state
\cite{Wan2011,Burkov2011,Bulmash2014} and in cold atom systems \cite{Xu2014}, as well as the experimental
identification of ``Dirac semimetals''
\cite{Neupane2013,Borisenko2013,Liu2014}, which have a pair of Weyl
nodes forced to overlap by time-reversal and inversion symmetry.
Although spectroscopic confirmation of a Weyl node in a real material
is still lacking, magnetotransport signatures 
consistent with Weyl nodes were reported for {\em BiSb} \cite{Kim2013}.

An important question that concerns the comparison of theory and 
experimental realizations is about the stability of the Weyl nodes
to the presence of disorder \cite{*[{The effect of a single impurity on the electronic structure of a Weyl node has been studied recently in\,}]  [{}] Huang2013}. This question is of particular fundamental
interest if the disorder is sufficiently smooth that scattering between 
different Weyl nodes is avoided, since disorder that does not satisfy this 
condition immediately removes any topological protection and leads to a 
trivial gapping of the spectrum and/or localization of the wavefunctions.

In the theoretical literature, the study of the effect of disorder on
a single Weyl node, without inclusion of electron-electron
interactions, goes back to the mid 1980s
\cite{Fradkin1986a,Fradkin1986}. Far away from the Weyl point the 
expected behavior resembles that of normal
metals: 
Disorder leads to diffusive dynamics, with a conductivity $\sigma$ that
decreases with increasing disorder strength. However, unlike a normal
metal, a Weyl semimetal has no transition into an Anderson-localized
phase in the limit of strong disorder \cite{Ryu2010}. 
Exactly at the Weyl point $\varepsilon=0$ a completely
different picture emerges: There is consensus that weak 
disorder is irrelevant
\cite{Fradkin1986a,Fradkin1986,Burkov2011a,Syzranov2014}, so that the
vanishing density of states $\nu(\varepsilon)\propto \varepsilon^{2}$ 
of the Hamiltonian (\ref{eq:H0}) is maintained at finite disorder strength
\cite{Biswas2014,Ominato2013}, up to possible rare-region effects \cite{Nandkishore2013}; 
For stronger disorder, a quantum phase transition
takes place, beyond which $\nu(0)$ is finite. There is no consensus 
for the implications of this scenario
for the conductivity $\sigma$, however. Using the self-consistent Born
approximation (SCBA), Ominato and Koshino find $\sigma = 0$ up to
the critical disorder strength, and a finite conductivity that
increases for stronger disorder \cite{Ominato2013},
whereas the Renormalization Group (RG) approach of Ref.\ \onlinecite{Syzranov2014}
gives a finite conductivity for subcritical disorder strengths. 
Boltzmann theory also gives a Weyl-point conductivity that is a decreasing
function of disorder strength, but there is no critical disorder strength
and $\sigma$ is finite throughout \cite{Burkov2011a,Hosur2012,Ominato2013}.

Remarkably, the question about the effect of disorder on a single Weyl
node has never been put to the test numerically. Recently, similar physics
has been investigated for a disordered Dirac semimetal employing diagonalization of a large tight
binding model \cite{Kobayashi2014}. The extension of these results to a Weyl semimetal
is problematic, however,
because any tight binding model with a Weyl node inevitably comes with 
its opposite-chirality partner
node \cite{Nielsen1981}, coupling to which cannot be fully avoided.
Yet, resorting to a numerical test is particularly relevant in the present
case, because none of the theoretical methods applied in the analytical
theory cited above are fully controlled at the Weyl point $\varepsilon=0$
(see Ref.\ \onlinecite{Syzranov2014} for a critical discussion).

In this Letter, we report numerical calculations of the transport
properties of a single Weyl node in the presence of a random
potential. We limit ourselves to transport at the Weyl point
$\varepsilon=0$, which is the energy at which the differences between
a Weyl semimetal and a normal metal are most pronounced. 
The focus on the nodal point is not entirely academic: In contrast to the
two-dimensional case (graphene or surface states of topological
insulators), where unintended doping generically
shifts the chemical potential away from the nodal point, in the bulk of three-dimensional Weyl semimetals $\varepsilon=0$ can be expected from stoichiometric filling of the
energy bands \cite{Biswas2014}.

Our results for the conductivity are qualitatively similar to the predictions
of the SCBA \cite{Ominato2013}, although
quantitatively the numerical results for the critical disorder
strength and for the conductivity approximately differ by a factor
two. In the weak-disorder phase the system is better characterized by
its conductance, which is finite, than by its conductivity, which is
zero within the accuracy of our calculations. A transport signature
that is nonzero in both phases is the Fano factor $F$, the ratio of the
shot-noise power and the conductance, which we show to be an excellent
indicator to discriminate between the pseudoballistic transport of the
weak-disorder phase and the diffusive transport of the strong-disorder
phase.

\emph{Model and numerical method.---}
Our numerical procedure closely follows Refs.\ 
\onlinecite{Bardarson2007,Adam2009}, which considered the effect of disorder
on the conductivity of graphene. We consider a Weyl semimetal of 
length $0 < x < L$ and transverse dimensions $0 < y,z < W$ with Hamiltonian
\begin{equation}
  H = H_{0}+U(\vr)\label{eq:H}
\end{equation}
where $U(\vr)$ is a Gaussian random potential with zero mean $\langle U_{\vq} 
\rangle = 0$ and fluctuations
\begin{equation}
  \langle U_{\vq} U_{\vq'}^* \rangle =
  \frac{\KK \xi \hbar^2 v^2}{W^2 L} e^{-q^2 \xi^2/2}\delta_{\vq,\vq'},
  \label{eq:UGauss}
\end{equation}
where $\xi$ is the correlation length and $\KK$ the dimensionless disorder 
strength. A similar random potential has been used in studies of the Dirac
equation in two dimensions \cite{Bardarson2007}.
For $x < 0$ and $x > L$ the Weyl semimetal is connected to ideal leads, which 
we model as Weyl semimetals with Hamiltonian $H_0 + V$, taking the 
limit $V \to -\infty$ \cite{Tworzydlo2006}. 
We numerically compute the transmission matrix $t$ at 
zero energy and determine the zero-temperature conductance using the Landauer 
formula $G(L,W) = (e^2/h) \mbox{tr}\, t t^{\dagger}$ and the Fano factor 
$F(L) = \mbox{tr}[t t^{\dagger}(1 - t t^{\dagger})]/\mbox{tr}\, 
t t^{\dagger}$. 
To quantize transverse momenta, we apply periodic or antiperiodic boundary 
conditions in the $y$ and $z$ directions, and truncate at 
$|q_{y}|,|q_{z}|\leq 2 M/\xi$, where we 
verified that the results do not depend on the cutoff $M$. To ensure bulk 
behavior, the width $W$ is taken large enough that the results do not 
depend on the boundary conditions and the scaling $G\propto W^{2}$, $F$ 
independent of $W$, holds. 

\noindent 
\begin{figure}
\noindent \begin{centering}
\includegraphics{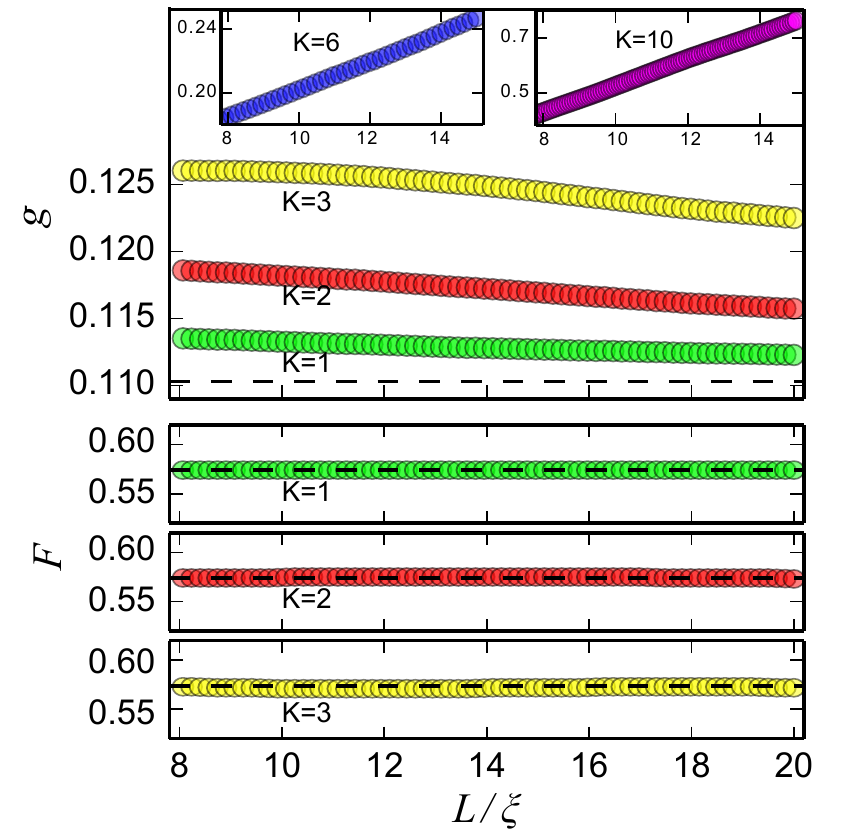}
\par\end{centering}

\caption{\label{fig:1} Dimensionless conductance $g$ referred to a cubic sample of size $L$ (top) and Fano factor $F$ (bottom) for a single Weyl cone with a random potential for disorder strengths $\KK = 1$, $2$, and $3$ in the pseudoballistic regime. The data represent a disorder average over at least $10$ realizations. The dashed lines refer to the clean limits $g_0$ and $F_0$ for an isotropic Weyl cone ($c=1$). For comparison, diffusive scaling of $g$ for $\KK=6, 10$ is shown in the insets.}
\end{figure}

\emph{Pseudoballistic regime.---} For the low-disorder regime, we rescale
the calculated conductance $G(L,W)$ to find the dimensionless conductance 
$g(L)$ of a cube with linear dimension $L$, 
\begin{equation}
  G(L,W) = \frac{e^2 W^2}{h L^2} g(L).
\end{equation}
In the absence of disorder $g$ and the Fano factor $F$ are independent of 
$L$ \cite{Baireuther2014}, taking the values
\begin{equation}
  g_0 = \frac{\ln 2}{2\pi} c,\ \ F_0 = \frac{1}{3} + \frac{1}{6 \ln 2}
  \approx 0.574,
\end{equation}
with $c$ a numerical factor that takes the value $c=1$ (so that $g_0 \approx
0.110$) for an isotropic Weyl cone. The results of numerical calculations of $g(L)$
and $F(L)$ for disorder strengths $\KK = 1$, $2$, and $3$ are
shown in Fig.\ \ref{fig:1}. The numerical data show that the presence of the
random potential $U(\vr)$ leads to a bulk conductance
$g$ that is always larger than the pseudoballistic value $g_0$, but also that the
conductance $g(L)$ is a bounded function of $L$ and monotonically
decreases in the large-$L$ limit. For the system sizes within
our reach this decrease is most
pronounced for weak disorder ($\KK = 1$), and less pronounced for
stronger disorder ($\KK = 3$), consistent with the theoretical
expectation that weak disorder is an irrelevant perturbation at
$\varepsilon=0$ \cite{Burkov2011a,Syzranov2014}. The fact that $g(L)$ 
remains bounded as a function of $L$ is consistent with a vanishing
conductivity $\sigma=0$. (A finite conductivity would correspond to
$g(L) \propto L$, see inset in Fig.\ \ref{fig:1}) The Fano factor $F$ takes the pseudoballistic
value $F_0$ for all system sizes considered. We 
postpone a further discussion of these results until the end of this
article.

\begin{figure}

\noindent \begin{centering}
\includegraphics{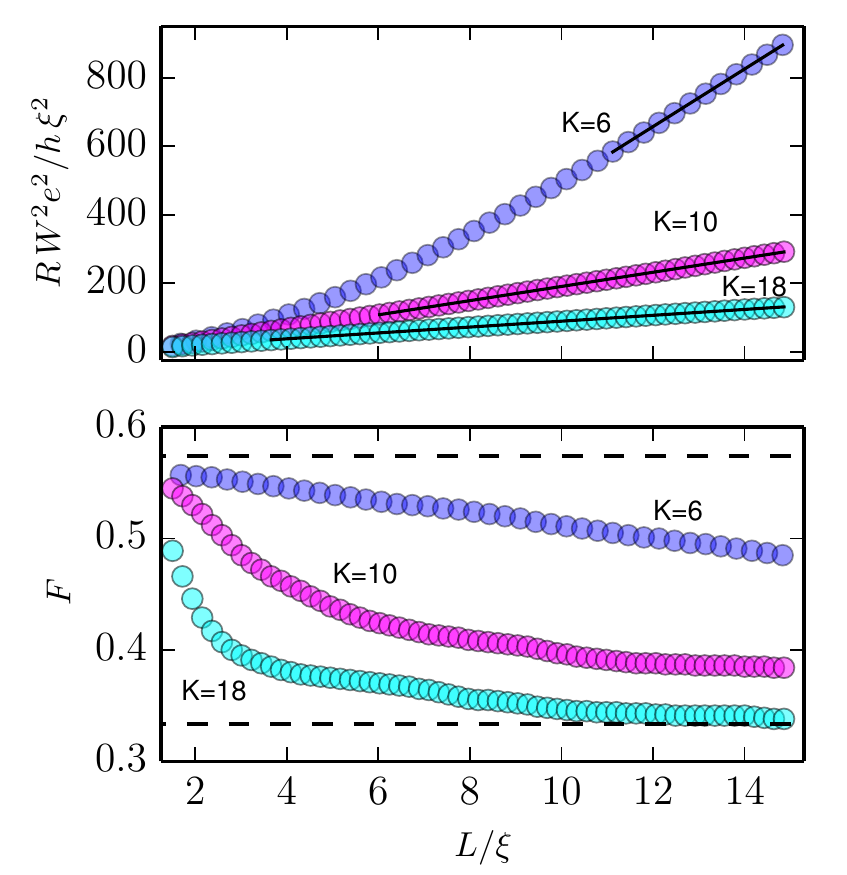}
\par\end{centering}

\caption{\label{fig:2} Resistance $R$ (top) and Fano factor $F$ (bottom) for a single Weyl cone vs.\ system length $L$, for disorder strengths $\KK = 6$, $10$, and $18$. The thin solid lines indicate the linear fit for the conductivity $\sigma$. The dashed lines refer to the pseudoballistic and diffusive limits for the Fano factor $F$. The data represent a disorder average over at least $100$ realizations.}
\end{figure}

\begin{figure}
\noindent \begin{centering}
\includegraphics{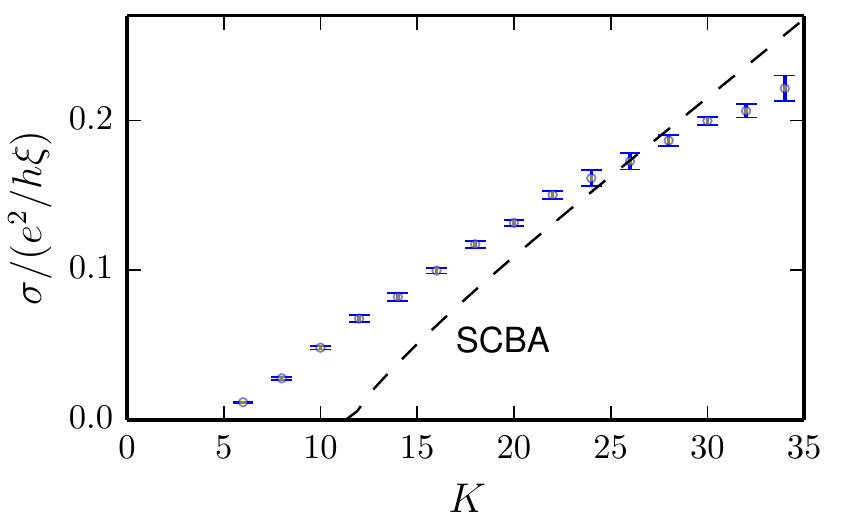}
\par\end{centering}

\caption{\label{fig:conductivity}Conductivity $\sigma$ for disordered Weyl
cone as a function of the disorder strength $\KK$. The data represent a 
disorder average over at least $50$ disorder realizations. The dashed line 
refers to the SCBA theory of Ref.\ \onlinecite{Ominato2013}. 
%The error bars 
%represent the uncertainty from mesoscopic fluctuations of the conductance as 
%well as from the uncertainty of fitting the slope in Fig.\ \ref{fig:2}.
}
\end{figure}

\emph{Diffusive regime.---} 
For stronger disorder, the conductivity $\sigma$ becomes finite.
Although $\sigma$ can in principle be obtained from the conductance using
the relation $G(L,W) = \sigma W^2/L$, we employ a slightly different
procedure to obtain $\sigma$ from the numerically calculated conductance
$G(L,W)$, in order to eliminate the effect of a finite contact resistance. 
Figure \ref{fig:2} shows the resistance $R(L,W) =
1/G(L,W)$ and the Fano factor $F(L)$ as a function of length $L$, for
disorder strengths $\KK=6$, $10$ and $18$. In the diffusive regime, one
expects $R(L,W) \propto L/W^2 \sigma$, so that the
conductivity can be calculated as $\sigma^{-1} = W^2 \partial R/\partial
L$. We indeed observe a linear $R$ vs.\ $L$ dependence for sufficiently
large $L$. The Fano factor
$F$ takes the diffusive value $F = 1/3$ for large $L$ for the stronger
disorder strengths such as $\KK=18$. For $\KK=6$ and $\KK=10$ the Fano factor $F$ is below the pseudoballistic limit and decreases with increasing $L$, but no limiting
value could be determined for the system sizes available in our calculations. The dependence of the conductivity $\sigma$ on
disorder strength $\KK$ is summarized in Fig.\ \ref{fig:conductivity}. 
We estimate that the conductivity is nonzero above a critical disorder
$\KKc \approx 5$, the behavior for $K$ just above $\KKc$ being consistent
with a linear increase $\propto \KK-\KKc$
\cite{Fradkin1986,Goswami2011,Ominato2013}; Finite-size effects prohibit a more
accurate determination of the critical disorder strength. Although we 
adopted the expression ``critical disorder strength'', we note that our
numerical analysis does not allow us to determine the precise 
nature of the transition. 
In passing, we also note that the conductance distribution is widest
around $\KKc$ (data not shown), a behavior well known from the
three-dimensional Anderson phase transition \cite{Slevin1997}.

\noindent 

A recent work by Ominato and Koshino \cite{Ominato2013} calculates the
Weyl-point conductivity $\sigma$ using the SCBA
but without further approximations, employing a correlated disorder potential 
compatible with the random potential used in the present numerical simulation. 
Relating the impurity model of Ref. \cite{Ominato2013} to our Gaussian model we
find a theoretical value $\KKc^{\rm SCBA} \approx 11.3$ and a conductivity
as shown by the dashed line Fig. \ref{fig:conductivity} \footnote{The theory
of Ref.\ \onlinecite{Ominato2013} can be applied to our calculations by 
making the substitutions $d_{0}\leftrightarrow\xi$ and $W\leftrightarrow \KK/2\pi$.}.
%The convention on Fourier transforms used in this work is $U(\mathbf{r})=\sum_{\mathbf{q}}e^{i\mathbf{q}\mathbf{r}}U_{\mathbf{q}}$, so that the relation between the Edwards model of Ref.\ \onlinecite{Ominato2013} and the Gaussian model used in this work is established as \cite{Akkermans}  $n_i u_0^2\leftrightarrow K_{0}\xi\left(\hbar v\right)^{2}/W^2L$.}}
Both the value of $\KKc^{\rm SCBA}$ and the slope of the SCBA conductivity
vs.\ disorder strength $\KK$ are roughly off by a factor of two from the numerical
results. 

\begin{figure}
\noindent \begin{centering}
\includegraphics{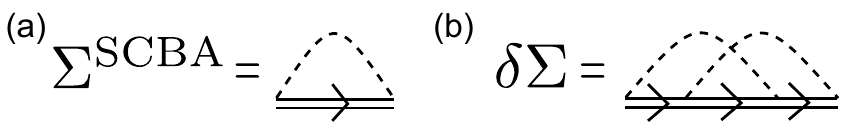}
\par\end{centering}

\caption{\label{fig:diagrams} Diagrammatic representation of the 
SCBA self energy $\Sigma^{\rm SCBA}$ (a) and the leading correction
$\delta \Sigma$ (b). The double solid lines denote the SCBA propagator;
dashed lines are disorder correlators.}
\end{figure}

In order to understand the quantitative failure of the SCBA we have analyzed
the corrections to the SCBA result for the self energy $\Sigma(\vk,\omega)$, 
which is related to the single-particle Green function ${\cal G}(\vk,\omega)$ 
through the standard relation ${\cal G}(\vk,\omega) = [\omega - H_0 - 
\Sigma(\vk,\omega)]^{-1}$. The diagrammatic expression
for $\Sigma(\vk,\omega)$ in the SCBA 
is shown in Fig.\ \ref{fig:diagrams}(a), where the 
double lines denote the single-particle 
Green function ${\cal G}$ with $\Sigma$ replaced by $\Sigma^{\rm SCBA}$.
Figure \ref{fig:diagrams}(b) contains the leading correction $\delta 
\Sigma$ to $\Sigma^{\rm SCBA}$. The consistency of the SCBA 
requires that $\delta\Sigma$ is parametrically smaller than $\Sigma^{\rm SCBA}$.
Indeed, for a standard disordered metal one finds $\delta\Sigma/\Sigma^{\rm SCBA}
={\cal O}(1/k_{\rm F}l)$ \cite{Rammer}, where $k_{\rm F}$ is the Fermi 
wavevector and $l$ the mean free path. 

For the Weyl semimetal at zero energy one has $k_{\rm F}=0$ and this
standard argument does not apply. We have calculated the leading
correction $\delta\Sigma$ at $k = 0$ and $\omega = 0$ using a
simplified model for the disorder potential \cite{Ominato2013}, in
which the Gaussian correlator (\ref{eq:UGauss}) is replaced by a
cutoff at $q = 2/\xi$,
\begin{equation}
  \langle U_{\vq} U_{\vq'}^* \rangle = \frac{\KK' \xi \hbar^2 v^2}{W^2 L}
  \Theta(2/\xi-q)
  \delta_{\vq,\vq'}.
  \label{eq:Usimpl}
\end{equation}
In this simplified model one has the critical disorder 
strength $\KKc' = \pi^2$ and the SCBA self energy 
$\Sigma(0,0)^{\rm SCBA} = (4 \pi i \hbar v/\xi) (1/\KK'-1/\KKc') 
\Theta(\KK' - \KKc')$ \cite{Ominato2013}. 
Calculation of the diagram of Fig.\ \ref{fig:diagrams}(b) for $\KK'$ close to
the critical disorder strength $\KKc'$ then gives \footnote{Details of
this calculation are given in the appendix.} %%%%% FOR PRL: supplemental material 
\begin{equation}
  \frac{\delta \Sigma(0,0)}{\Sigma(0,0)^{\rm SCBA}} \simeq
  0.62 + 11 \left( \frac{1}{\KKc'}-\frac{1}{\KK'} \right),
  \label{eq:estimate}
\end{equation}
which is not parametrically small. Since the simplified model (\ref{eq:Usimpl})
does not qualitatively differ from the Gaussian model used in the numerical
calculations \cite{Ominato2013}, we expect that this result carries over to that 
case, too.

\noindent 

\emph{Discussion.---}
In the framework of Drude transport
theory for normal metals, the quasiparticles at the Fermi energy are
endowed with a mean free path, which becomes shorter if the
disorder becomes stronger. At the same time, the presence of a random 
impurity potential has negligible effect on the density of states. The
result is a conductivity that decreases upon increasing the disorder
strength. 
In contrast, for a Weyl node at the degeneracy point it is the disorder
which generates the density of states 
\cite{Fradkin1986,Fradkin1986a,Burkov2011a,Biswas2014,Ominato2013,Syzranov2014}, a finite density of states appearing
only above a certain critical disorder strength. As a result of this
vastly different physical mechanism, a Weyl node at the degeneracy point
shows behavior opposite to that of a normal metal: Increasing disorder beyond the critical disorder strength leads to an increase of the conductivity. This remarkable theoretical prediction has been confirmed in our
numerical calculations.

The increase in conductivity with disorder is reminiscent of the 
two-dimensional Dirac Hamiltonian $H_{0}^{2d}\propto v(k_{x}\sigma_{x}+
k_{y}\sigma_{y})$, for which the conductivity $\sigma$ was also found
to be an increasing function of disorder strength 
\cite{Bardarson2007,Adam2009,DasSarma2012}. A fundamental difference
with $H_{0}^{2d}$ is, however, that $H_{0}^{2d}$ has a finite conductivity for 
all disorder strengths, whereas the Weyl semimetal at the degeneracy
point requires a minimum disorder strength for diffusive behavior to
set in.

For the two-dimensional Dirac Hamiltonian, the inverted dependence of
conductivity on disorder strength was found to be related to the fact 
that $H_{0}^{2d}$ (with a disorder term but without the condition that 
the disorder be smooth, because of the absence of other Dirac 
nodes) is the surface theory of a three-dimensional 
time-reversal invariant topological insulator \cite{Ryu2010}.
Similarly, the surface theory of a hypothetical four-dimensional topological
insulator is described by the Hamiltonian $H_{0}$ of Eq. (\ref{eq:H0}). 
Thus, it is expected on general grounds that 
$H_{0}$ evades localization \cite{Ryu2010}. Our numerical results are
consistent with this expectation. Indeed, although the conductivity
$\sigma$ vanishes in the weak-disorder regime, the conductance $g$
remains finite. It is a finite conductance, not a finite conductivity, 
which is the proper signature of absence of localization \cite{Imry2002}.

There is a subtle but important difference between the 
numerical calculations we performed here and the analytical theories
of the conductivity cited in the introduction: In our calculations, the 
conductivity $\sigma$
is obtained from the conductance $G$ of a finite-size sample, for which
the energy $\varepsilon$ is set to zero at the beginning of the calculation.
In contrast, in the RG, SCBA, and Boltzmann theories, the sample size is
infinite and the limit $\varepsilon 
\to 0$ is taken at the end of the calculation 
\cite{Burkov2011a,Syzranov2014,Biswas2014,Ominato2013}. This different
order of limits may be responsible for the qualitative difference with
Refs.\ \onlinecite{Burkov2011a,Syzranov2014,Biswas2014}, which predict
a finite conductivity in the limit $\varepsilon \to 0$. Which order of
limits is relevant for experiments depends on the competition between the
finite sample size $L$ and the finite temperature or doping 
\cite{Syzranov2014,Biswas2014} --- although the latter is expected to 
be intrinsically small. The order of the limits
$\varepsilon \to 0$ and $L \to \infty$ does not affect the comparison 
to the SCBA, because this theory predicts $\sigma = 0$ even if the limit
$\varepsilon \to 0$ is taken at the end of the calculation \cite{Ominato2013}.
Above the critical disorder strength, the self energy at $\varepsilon=0$ 
acquires a nonzero (imaginary) value and the order of limits issue is 
no longer relevant.

Our numerical calculations have shown that the conductance
$g$ and the Fano factor $F$ contain important additional information that is not
contained in the conductivity $\sigma$. This is particularly relevant
for the pseudoballistic weak disorder regime, where $\sigma$
vanishes, whereas $g$ and $F$ take on
nonzero values. A three-dimensional phase with a finite 
scale-independent bulk conductance is known from the Anderson metal-insulator
transition, where it occurs at the critical disorder strength. A crucial
difference of the pseudoballistic phase at the Weyl point is that its
scale-independent conductance represents an attractive fixed point, which
requires no fine tuning of disorder strength.

%For experiments, where disorder is of course inevitably present, our
%refinement of the critical disorder strength can be crucial to interpret
%transport and spectroscopic data that will strongly depend on the
%actual disorder being larger or smaller than the critical disorder
%strength. However, for weak disorder, electron-electron interactions
%are expected to dominate transport quantities in Weyl semimetals.
%This can be seen from a simple power counting RG procedure\cite{Burkov2011a},
%where interactions are marginally relevant and are expected to considerably
%change the nature of the ballistic fixed point.

\emph{Acknowledgments.---} We acknowledge helpful discussions with
Georg Schwiete, Martin Schneider, Sergey Syzranov, Leo Radzihovsky, Victor Gurarie and we thank J\"org Behrmann for support
with the computations. Financial support was granted by the Helmholtz
Virtual Institute ``New states of matter and their excitations'',
by the Alexander von Humboldt Foundation in the framework of the Alexander
von Humboldt Professorship, endowed by the Federal Ministry of Education
and Research, and by DFG\textquoteright{}s Emmy Noether program (BE
5233/1-1).

\bibliography{weyl}

\begin{thebibliography}{46}%
\makeatletter
\providecommand \@ifxundefined [1]{%
 \@ifx{#1\undefined}
}%
\providecommand \@ifnum [1]{%
 \ifnum #1\expandafter \@firstoftwo
 \else \expandafter \@secondoftwo
 \fi
}%
\providecommand \@ifx [1]{%
 \ifx #1\expandafter \@firstoftwo
 \else \expandafter \@secondoftwo
 \fi
}%
\providecommand \natexlab [1]{#1}%
\providecommand \enquote  [1]{``#1''}%
\providecommand \bibnamefont  [1]{#1}%
\providecommand \bibfnamefont [1]{#1}%
\providecommand \citenamefont [1]{#1}%
\providecommand \href@noop [0]{\@secondoftwo}%
\providecommand \href [0]{\begingroup \@sanitize@url \@href}%
\providecommand \@href[1]{\@@startlink{#1}\@@href}%
\providecommand \@@href[1]{\endgroup#1\@@endlink}%
\providecommand \@sanitize@url [0]{\catcode `\\12\catcode `\$12\catcode
  `\&12\catcode `\#12\catcode `\^12\catcode `\_12\catcode `\%12\relax}%
\providecommand \@@startlink[1]{}%
\providecommand \@@endlink[0]{}%
\providecommand \url  [0]{\begingroup\@sanitize@url \@url }%
\providecommand \@url [1]{\endgroup\@href {#1}{\urlprefix }}%
\providecommand \urlprefix  [0]{URL }%
\providecommand \Eprint [0]{\href }%
\providecommand \doibase [0]{http://dx.doi.org/}%
\providecommand \selectlanguage [0]{\@gobble}%
\providecommand \bibinfo  [0]{\@secondoftwo}%
\providecommand \bibfield  [0]{\@secondoftwo}%
\providecommand \translation [1]{[#1]}%
\providecommand \BibitemOpen [0]{}%
\providecommand \bibitemStop [0]{}%
\providecommand \bibitemNoStop [0]{.\EOS\space}%
\providecommand \EOS [0]{\spacefactor3000\relax}%
\providecommand \BibitemShut  [1]{\csname bibitem#1\endcsname}%
\let\auto@bib@innerbib\@empty
%</preamble>
\bibitem [{\citenamefont {Hasan}\ and\ \citenamefont {Kane}(2010)}]{Hasan2010}%
  \BibitemOpen
  \bibfield  {author} {\bibinfo {author} {\bibfnamefont {M.~Z.}\ \bibnamefont
  {Hasan}}\ and\ \bibinfo {author} {\bibfnamefont {C.~L.}\ \bibnamefont
  {Kane}},\ }\href {\doibase 10.1103/RevModPhys.82.3045} {\bibfield  {journal}
  {\bibinfo  {journal} {Rev. Mod. Phys.}\ }\textbf {\bibinfo {volume} {82}},\
  \bibinfo {pages} {3045} (\bibinfo {year} {2010})}\BibitemShut {NoStop}%
\bibitem [{\citenamefont {Bernevig}(2013)}]{Bernevig2013}%
  \BibitemOpen
  \bibfield  {author} {\bibinfo {author} {\bibfnamefont {B.~A.}\ \bibnamefont
  {Bernevig}},\ }\href@noop {} {\emph {\bibinfo {title} {{Topological
  Insulators and Topological Superconductors}}}}\ (\bibinfo  {publisher}
  {Princeton University Press},\ \bibinfo {year} {2013})\BibitemShut {NoStop}%
\bibitem [{\citenamefont {Wen}\ and\ \citenamefont {Zee}(2002)}]{Wen2002}%
  \BibitemOpen
  \bibfield  {author} {\bibinfo {author} {\bibfnamefont {X.~G.}\ \bibnamefont
  {Wen}}\ and\ \bibinfo {author} {\bibfnamefont {A.}~\bibnamefont {Zee}},\
  }\href {\doibase 10.1103/PhysRevB.66.235110} {\bibfield  {journal} {\bibinfo
  {journal} {Phys. Rev. B}\ }\textbf {\bibinfo {volume} {66}},\ \bibinfo
  {pages} {235110} (\bibinfo {year} {2002})}\BibitemShut {NoStop}%
\bibitem [{\citenamefont {Ho\ifmmode~\check{r}\else
  \v{r}\fi{}ava}(2005)}]{Horava2005}%
  \BibitemOpen
  \bibfield  {author} {\bibinfo {author} {\bibfnamefont {P.}~\bibnamefont
  {Ho\ifmmode~\check{r}\else \v{r}\fi{}ava}},\ }\href {\doibase
  10.1103/PhysRevLett.95.016405} {\bibfield  {journal} {\bibinfo  {journal}
  {Phys. Rev. Lett.}\ }\textbf {\bibinfo {volume} {95}},\ \bibinfo {pages}
  {016405} (\bibinfo {year} {2005})}\BibitemShut {NoStop}%
\bibitem [{\citenamefont {Sato}(2006)}]{Sato2006}%
  \BibitemOpen
  \bibfield  {author} {\bibinfo {author} {\bibfnamefont {M.}~\bibnamefont
  {Sato}},\ }\href {\doibase 10.1103/PhysRevB.73.214502} {\bibfield  {journal}
  {\bibinfo  {journal} {Phys. Rev. B}\ }\textbf {\bibinfo {volume} {73}},\
  \bibinfo {pages} {214502} (\bibinfo {year} {2006})}\BibitemShut {NoStop}%
\bibitem [{\citenamefont {B\'eri}(2010)}]{Beri2010}%
  \BibitemOpen
  \bibfield  {author} {\bibinfo {author} {\bibfnamefont {B.}~\bibnamefont
  {B\'eri}},\ }\href {\doibase 10.1103/PhysRevB.81.134515} {\bibfield
  {journal} {\bibinfo  {journal} {Phys. Rev. B}\ }\textbf {\bibinfo {volume}
  {81}},\ \bibinfo {pages} {134515} (\bibinfo {year} {2010})}\BibitemShut
  {NoStop}%
\bibitem [{\citenamefont {Volovik}(2011)}]{Volovik2011}%
  \BibitemOpen
  \bibfield  {author} {\bibinfo {author} {\bibfnamefont {G.}~\bibnamefont
  {Volovik}},\ }\href {\doibase 10.1134/S0021364011020147} {\bibfield
  {journal} {\bibinfo  {journal} {JETP Letters}\ }\textbf {\bibinfo {volume}
  {93}},\ \bibinfo {pages} {66} (\bibinfo {year} {2011})}\BibitemShut {NoStop}%
\bibitem [{\citenamefont {Bernard}\ \emph {et~al.}(2012)\citenamefont
  {Bernard}, \citenamefont {Kim},\ and\ \citenamefont {LeClair}}]{Bernard2012}%
  \BibitemOpen
  \bibfield  {author} {\bibinfo {author} {\bibfnamefont {D.}~\bibnamefont
  {Bernard}}, \bibinfo {author} {\bibfnamefont {E.-A.}\ \bibnamefont {Kim}}, \
  and\ \bibinfo {author} {\bibfnamefont {A.}~\bibnamefont {LeClair}},\ }\href
  {\doibase 10.1103/PhysRevB.86.205116} {\bibfield  {journal} {\bibinfo
  {journal} {Phys. Rev. B}\ }\textbf {\bibinfo {volume} {86}},\ \bibinfo
  {pages} {205116} (\bibinfo {year} {2012})}\BibitemShut {NoStop}%
\bibitem [{\citenamefont {Zhao}\ and\ \citenamefont {Wang}(2013)}]{Zhao2013}%
  \BibitemOpen
  \bibfield  {author} {\bibinfo {author} {\bibfnamefont {Y.~X.}\ \bibnamefont
  {Zhao}}\ and\ \bibinfo {author} {\bibfnamefont {Z.~D.}\ \bibnamefont
  {Wang}},\ }\href {\doibase 10.1103/PhysRevLett.110.240404} {\bibfield
  {journal} {\bibinfo  {journal} {Phys. Rev. Lett.}\ }\textbf {\bibinfo
  {volume} {110}},\ \bibinfo {pages} {240404} (\bibinfo {year}
  {2013})}\BibitemShut {NoStop}%
\bibitem [{\citenamefont {Matsuura}\ \emph {et~al.}(2013)\citenamefont
  {Matsuura}, \citenamefont {Chang}, \citenamefont {Schnyder},\ and\
  \citenamefont {Ryu}}]{Matsuura2013}%
  \BibitemOpen
  \bibfield  {author} {\bibinfo {author} {\bibfnamefont {S.}~\bibnamefont
  {Matsuura}}, \bibinfo {author} {\bibfnamefont {P.-Y.}\ \bibnamefont {Chang}},
  \bibinfo {author} {\bibfnamefont {A.~P.}\ \bibnamefont {Schnyder}}, \ and\
  \bibinfo {author} {\bibfnamefont {S.}~\bibnamefont {Ryu}},\ }\href {\doibase
  10.1088/1367-2630/15/6/065001} {\bibfield  {journal} {\bibinfo  {journal}
  {New J. Phys.}\ }\textbf {\bibinfo {volume} {15}},\ \bibinfo {pages} {065001}
  (\bibinfo {year} {2013})}\BibitemShut {NoStop}%
\bibitem [{\citenamefont {Wallace}(1947)}]{Wallace1947}%
  \BibitemOpen
  \bibfield  {author} {\bibinfo {author} {\bibfnamefont {P.~R.}\ \bibnamefont
  {Wallace}},\ }\href@noop {} {\bibfield  {journal} {\bibinfo  {journal} {Phys.
  Rev.}\ }\textbf {\bibinfo {volume} {71}},\ \bibinfo {pages} {622} (\bibinfo
  {year} {1947})}\BibitemShut {NoStop}%
\bibitem [{\citenamefont {Murakami}(2007)}]{Murakami2007}%
  \BibitemOpen
  \bibfield  {author} {\bibinfo {author} {\bibfnamefont {S.}~\bibnamefont
  {Murakami}},\ }\href@noop {} {\bibfield  {journal} {\bibinfo  {journal} {New
  J. Phys.}\ }\textbf {\bibinfo {volume} {356}} (\bibinfo {year}
  {2007})}\BibitemShut {NoStop}%
\bibitem [{\citenamefont {Wan}\ \emph {et~al.}(2011)\citenamefont {Wan},
  \citenamefont {Turner}, \citenamefont {Vishwanath},\ and\ \citenamefont
  {Savrasov}}]{Wan2011}%
  \BibitemOpen
  \bibfield  {author} {\bibinfo {author} {\bibfnamefont {X.}~\bibnamefont
  {Wan}}, \bibinfo {author} {\bibfnamefont {A.~M.}\ \bibnamefont {Turner}},
  \bibinfo {author} {\bibfnamefont {A.}~\bibnamefont {Vishwanath}}, \ and\
  \bibinfo {author} {\bibfnamefont {S.~Y.}\ \bibnamefont {Savrasov}},\ }\href
  {\doibase 10.1103/PhysRevB.83.205101} {\bibfield  {journal} {\bibinfo
  {journal} {Phys. Rev. B}\ }\textbf {\bibinfo {volume} {83}},\ \bibinfo
  {pages} {205101} (\bibinfo {year} {2011})}\BibitemShut {NoStop}%
\bibitem [{\citenamefont {Hosur}\ and\ \citenamefont {Qi}(2013)}]{Hosur2013}%
  \BibitemOpen
  \bibfield  {author} {\bibinfo {author} {\bibfnamefont {P.}~\bibnamefont
  {Hosur}}\ and\ \bibinfo {author} {\bibfnamefont {X.}~\bibnamefont {Qi}},\
  }\href@noop {} {\bibfield  {journal} {\bibinfo  {journal} {C. R. Physique}\
  }\textbf {\bibinfo {volume} {14}},\ \bibinfo {pages} {857} (\bibinfo {year}
  {2013})}\BibitemShut {NoStop}%
\bibitem [{\citenamefont {Goswami}\ and\ \citenamefont
  {Tewari}(2013)}]{Goswami2013}%
  \BibitemOpen
  \bibfield  {author} {\bibinfo {author} {\bibfnamefont {P.}~\bibnamefont
  {Goswami}}\ and\ \bibinfo {author} {\bibfnamefont {S.}~\bibnamefont
  {Tewari}},\ }\href {\doibase 10.1103/PhysRevB.88.245107} {\bibfield
  {journal} {\bibinfo  {journal} {Phys. Rev. B}\ }\textbf {\bibinfo {volume}
  {88}},\ \bibinfo {pages} {245107} (\bibinfo {year} {2013})}\BibitemShut
  {NoStop}%
\bibitem [{\citenamefont {Vazifeh}\ and\ \citenamefont
  {Franz}(2013)}]{Vazifeh2013}%
  \BibitemOpen
  \bibfield  {author} {\bibinfo {author} {\bibfnamefont {M.~M.}\ \bibnamefont
  {Vazifeh}}\ and\ \bibinfo {author} {\bibfnamefont {M.}~\bibnamefont
  {Franz}},\ }\href {\doibase 10.1103/PhysRevLett.111.027201} {\bibfield
  {journal} {\bibinfo  {journal} {Phys. Rev. Lett.}\ }\textbf {\bibinfo
  {volume} {111}},\ \bibinfo {pages} {027201} (\bibinfo {year}
  {2013})}\BibitemShut {NoStop}%
\bibitem [{\citenamefont {Baireuther}\ \emph {et~al.}(2014)\citenamefont
  {Baireuther}, \citenamefont {Edge}, \citenamefont {Fulga}, \citenamefont
  {Beenakker},\ and\ \citenamefont {Tworzydlo}}]{Baireuther2014}%
  \BibitemOpen
  \bibfield  {author} {\bibinfo {author} {\bibfnamefont {P.}~\bibnamefont
  {Baireuther}}, \bibinfo {author} {\bibfnamefont {J.~M.}\ \bibnamefont
  {Edge}}, \bibinfo {author} {\bibfnamefont {I.~C.}\ \bibnamefont {Fulga}},
  \bibinfo {author} {\bibfnamefont {C.~W.~J.}\ \bibnamefont {Beenakker}}, \
  and\ \bibinfo {author} {\bibfnamefont {J.}~\bibnamefont {Tworzydlo}},\ }\href
  {\doibase 10.1103/PhysRevB.89.035410} {\bibfield  {journal} {\bibinfo
  {journal} {Phys. Rev. B}\ }\textbf {\bibinfo {volume} {89}},\ \bibinfo
  {pages} {035410} (\bibinfo {year} {2014})}\BibitemShut {NoStop}%
\bibitem [{\citenamefont {Burkov}\ and\ \citenamefont
  {Balents}(2011)}]{Burkov2011}%
  \BibitemOpen
  \bibfield  {author} {\bibinfo {author} {\bibfnamefont {A.~A.}\ \bibnamefont
  {Burkov}}\ and\ \bibinfo {author} {\bibfnamefont {L.}~\bibnamefont
  {Balents}},\ }\href {\doibase 10.1103/PhysRevLett.107.127205} {\bibfield
  {journal} {\bibinfo  {journal} {Phys. Rev. Lett.}\ }\textbf {\bibinfo
  {volume} {107}},\ \bibinfo {pages} {127205} (\bibinfo {year}
  {2011})}\BibitemShut {NoStop}%
\bibitem [{\citenamefont {Bulmash}\ \emph {et~al.}(2014)\citenamefont
  {Bulmash}, \citenamefont {Liu},\ and\ \citenamefont {Qi}}]{Bulmash2014}%
  \BibitemOpen
  \bibfield  {author} {\bibinfo {author} {\bibfnamefont {D.}~\bibnamefont
  {Bulmash}}, \bibinfo {author} {\bibfnamefont {C.-X.}\ \bibnamefont {Liu}}, \
  and\ \bibinfo {author} {\bibfnamefont {X.-L.}\ \bibnamefont {Qi}},\
  }\href@noop {} {\bibfield  {journal} {\bibinfo  {journal} {Phys. Rev. B}\
  }\textbf {\bibinfo {volume} {89}} (\bibinfo {year} {2014})}\BibitemShut
  {NoStop}%
\bibitem [{\citenamefont {Xu}\ \emph {et~al.}(2014)\citenamefont {Xu},
  \citenamefont {Chu},\ and\ \citenamefont {Zhang}}]{Xu2014}%
  \BibitemOpen
  \bibfield  {author} {\bibinfo {author} {\bibfnamefont {Y.}~\bibnamefont
  {Xu}}, \bibinfo {author} {\bibfnamefont {R.-L.}\ \bibnamefont {Chu}}, \ and\
  \bibinfo {author} {\bibfnamefont {C.}~\bibnamefont {Zhang}},\ }\href@noop {}
  {\bibfield  {journal} {\bibinfo  {journal} {Phys. Rev. Lett.}\ }\textbf
  {\bibinfo {volume} {112}},\ \bibinfo {pages} {136402} (\bibinfo {year}
  {2014})}\BibitemShut {NoStop}%
\bibitem [{\citenamefont {Neupane}\ \emph {et~al.}(2014)\citenamefont
  {Neupane}, \citenamefont {Xu}, \citenamefont {Sankar}, \citenamefont
  {Alidoust}, \citenamefont {Bian}, \citenamefont {Liu}, \citenamefont
  {Belopolski}, \citenamefont {Chang}, \citenamefont {Jeng}, \citenamefont
  {Lin}, \citenamefont {Bansil}, \citenamefont {Chou},\ and\ \citenamefont
  {Hasan}}]{Neupane2013}%
  \BibitemOpen
  \bibfield  {author} {\bibinfo {author} {\bibfnamefont {M.}~\bibnamefont
  {Neupane}}, \bibinfo {author} {\bibfnamefont {S.}~\bibnamefont {Xu}},
  \bibinfo {author} {\bibfnamefont {R.}~\bibnamefont {Sankar}}, \bibinfo
  {author} {\bibfnamefont {N.}~\bibnamefont {Alidoust}}, \bibinfo {author}
  {\bibfnamefont {G.}~\bibnamefont {Bian}}, \bibinfo {author} {\bibfnamefont
  {C.}~\bibnamefont {Liu}}, \bibinfo {author} {\bibfnamefont {I.}~\bibnamefont
  {Belopolski}}, \bibinfo {author} {\bibfnamefont {T.}~\bibnamefont {Chang}},
  \bibinfo {author} {\bibfnamefont {H.}~\bibnamefont {Jeng}}, \bibinfo {author}
  {\bibfnamefont {H.}~\bibnamefont {Lin}}, \bibinfo {author} {\bibfnamefont
  {A.}~\bibnamefont {Bansil}}, \bibinfo {author} {\bibfnamefont
  {F.}~\bibnamefont {Chou}}, \ and\ \bibinfo {author} {\bibfnamefont {M.~Z.}\
  \bibnamefont {Hasan}},\ }\href@noop {} {\bibfield  {journal} {\bibinfo
  {journal} {Nature Communications}\ }\textbf {\bibinfo {volume} {5:3786}}
  (\bibinfo {year} {2014})}\BibitemShut {NoStop}%
\bibitem [{\citenamefont {Borisenko}\ and\ \citenamefont
  {Gibson}(2013)}]{Borisenko2013}%
  \BibitemOpen
  \bibfield  {author} {\bibinfo {author} {\bibfnamefont {S.}~\bibnamefont
  {Borisenko}}\ and\ \bibinfo {author} {\bibfnamefont {Q.}~\bibnamefont
  {Gibson}},\ }\href@noop {} {\bibfield  {journal} {\bibinfo  {journal}
  {arXiv:1309.7978v1}\ } (\bibinfo {year} {2013})}\BibitemShut {NoStop}%
\bibitem [{\citenamefont {Liu}\ \emph {et~al.}(2014)\citenamefont {Liu},
  \citenamefont {Zhou}, \citenamefont {Zhang}, \citenamefont {Wang},
  \citenamefont {Weng}, \citenamefont {Prabhakaran}, \citenamefont {Mo},
  \citenamefont {Shen}, \citenamefont {Fang}, \citenamefont {Dai},
  \citenamefont {Hussain},\ and\ \citenamefont {Chen}}]{Liu2014}%
  \BibitemOpen
  \bibfield  {author} {\bibinfo {author} {\bibfnamefont {Z.~K.}\ \bibnamefont
  {Liu}}, \bibinfo {author} {\bibfnamefont {B.}~\bibnamefont {Zhou}}, \bibinfo
  {author} {\bibfnamefont {Y.}~\bibnamefont {Zhang}}, \bibinfo {author}
  {\bibfnamefont {Z.~J.}\ \bibnamefont {Wang}}, \bibinfo {author}
  {\bibfnamefont {H.~M.}\ \bibnamefont {Weng}}, \bibinfo {author}
  {\bibfnamefont {D.}~\bibnamefont {Prabhakaran}}, \bibinfo {author}
  {\bibfnamefont {S.-K.}\ \bibnamefont {Mo}}, \bibinfo {author} {\bibfnamefont
  {Z.~X.}\ \bibnamefont {Shen}}, \bibinfo {author} {\bibfnamefont
  {Z.}~\bibnamefont {Fang}}, \bibinfo {author} {\bibfnamefont {X.}~\bibnamefont
  {Dai}}, \bibinfo {author} {\bibfnamefont {Z.}~\bibnamefont {Hussain}}, \ and\
  \bibinfo {author} {\bibfnamefont {Y.~L.}\ \bibnamefont {Chen}},\ }\href
  {\doibase 10.1126/science.1245085} {\bibfield  {journal} {\bibinfo  {journal}
  {Science}\ }\textbf {\bibinfo {volume} {343}},\ \bibinfo {pages} {864}
  (\bibinfo {year} {2014})}\BibitemShut {NoStop}%
\bibitem [{\citenamefont {Kim}\ \emph {et~al.}(2013)\citenamefont {Kim},
  \citenamefont {Kim}, \citenamefont {Wang}, \citenamefont {Sasaki},
  \citenamefont {Satoh}, \citenamefont {Ohnishi}, \citenamefont {Kitaura},
  \citenamefont {Yang},\ and\ \citenamefont {Li}}]{Kim2013}%
  \BibitemOpen
  \bibfield  {author} {\bibinfo {author} {\bibfnamefont {H.-J.}\ \bibnamefont
  {Kim}}, \bibinfo {author} {\bibfnamefont {K.-S.}\ \bibnamefont {Kim}},
  \bibinfo {author} {\bibfnamefont {J.-F.}\ \bibnamefont {Wang}}, \bibinfo
  {author} {\bibfnamefont {M.}~\bibnamefont {Sasaki}}, \bibinfo {author}
  {\bibfnamefont {N.}~\bibnamefont {Satoh}}, \bibinfo {author} {\bibfnamefont
  {A.}~\bibnamefont {Ohnishi}}, \bibinfo {author} {\bibfnamefont
  {M.}~\bibnamefont {Kitaura}}, \bibinfo {author} {\bibfnamefont
  {M.}~\bibnamefont {Yang}}, \ and\ \bibinfo {author} {\bibfnamefont
  {L.}~\bibnamefont {Li}},\ }\href {\doibase 10.1103/PhysRevLett.111.246603}
  {\bibfield  {journal} {\bibinfo  {journal} {Phys. Rev. Lett.}\ }\textbf
  {\bibinfo {volume} {111}},\ \bibinfo {pages} {246603} (\bibinfo {year}
  {2013})}\BibitemShut {NoStop}%
\bibitem [{\citenamefont {Huang}\ \emph {et~al.}(2013)\citenamefont {Huang},
  \citenamefont {Das}, \citenamefont {Balatsky},\ and\ \citenamefont
  {Arovas}}]{Huang2013}%
  \BibitemOpen
  \bibfield  {author} {\bibinfo {author} {\bibfnamefont {Z.}~\bibnamefont
  {Huang}}, \bibinfo {author} {\bibfnamefont {T.}~\bibnamefont {Das}}, \bibinfo
  {author} {\bibfnamefont {A.}~\bibnamefont {Balatsky}}, \ and\ \bibinfo
  {author} {\bibfnamefont {D.}~\bibnamefont {Arovas}},\ }\href {\doibase
  10.1103/PhysRevB.87.155123} {\bibfield  {journal} {\bibinfo  {journal} {Phys.
  Rev. B}\ }\textbf {\bibinfo {volume} {87}},\ \bibinfo {pages} {155123}
  (\bibinfo {year} {2013})}\BibitemShut {NoStop}%
\bibitem [{\citenamefont {Fradkin}(1986{\natexlab{a}})}]{Fradkin1986a}%
  \BibitemOpen
  \bibfield  {author} {\bibinfo {author} {\bibfnamefont {E.}~\bibnamefont
  {Fradkin}},\ }\href@noop {} {\bibfield  {journal} {\bibinfo  {journal} {Phys.
  Rev. B}\ }\textbf {\bibinfo {volume} {33}},\ \bibinfo {pages} {3257}
  (\bibinfo {year} {1986}{\natexlab{a}})}\BibitemShut {NoStop}%
\bibitem [{\citenamefont {Fradkin}(1986{\natexlab{b}})}]{Fradkin1986}%
  \BibitemOpen
  \bibfield  {author} {\bibinfo {author} {\bibfnamefont {E.}~\bibnamefont
  {Fradkin}},\ }\href@noop {} {\bibfield  {journal} {\bibinfo  {journal} {Phys.
  Rev. B}\ }\textbf {\bibinfo {volume} {33}},\ \bibinfo {pages} {3263}
  (\bibinfo {year} {1986}{\natexlab{b}})}\BibitemShut {NoStop}%
\bibitem [{\citenamefont {Ryu}\ \emph {et~al.}(2010)\citenamefont {Ryu},
  \citenamefont {Schnyder}, \citenamefont {Furusaki},\ and\ \citenamefont
  {Ludwig}}]{Ryu2010}%
  \BibitemOpen
  \bibfield  {author} {\bibinfo {author} {\bibfnamefont {S.}~\bibnamefont
  {Ryu}}, \bibinfo {author} {\bibfnamefont {A.~P.}\ \bibnamefont {Schnyder}},
  \bibinfo {author} {\bibfnamefont {A.}~\bibnamefont {Furusaki}}, \ and\
  \bibinfo {author} {\bibfnamefont {A.~W.~W.}\ \bibnamefont {Ludwig}},\ }\href
  {\doibase 10.1088/1367-2630/12/6/065010} {\bibfield  {journal} {\bibinfo
  {journal} {New J. Phys.}\ }\textbf {\bibinfo {volume} {12}},\ \bibinfo
  {pages} {065010} (\bibinfo {year} {2010})}\BibitemShut {NoStop}%
\bibitem [{\citenamefont {Burkov}\ \emph {et~al.}(2011)\citenamefont {Burkov},
  \citenamefont {Hook},\ and\ \citenamefont {Balents}}]{Burkov2011a}%
  \BibitemOpen
  \bibfield  {author} {\bibinfo {author} {\bibfnamefont {A.~A.}\ \bibnamefont
  {Burkov}}, \bibinfo {author} {\bibfnamefont {M.~D.}\ \bibnamefont {Hook}}, \
  and\ \bibinfo {author} {\bibfnamefont {L.}~\bibnamefont {Balents}},\ }\href
  {\doibase 10.1103/PhysRevB.84.235126} {\bibfield  {journal} {\bibinfo
  {journal} {Phys. Rev. B}\ }\textbf {\bibinfo {volume} {84}},\ \bibinfo
  {pages} {235126} (\bibinfo {year} {2011})}\BibitemShut {NoStop}%
\bibitem [{\citenamefont {Syzranov}\ \emph {et~al.}(2014)\citenamefont
  {Syzranov}, \citenamefont {Radzihovsky},\ and\ \citenamefont
  {Gurarie}}]{Syzranov2014}%
  \BibitemOpen
  \bibfield  {author} {\bibinfo {author} {\bibfnamefont {S.~V.}\ \bibnamefont
  {Syzranov}}, \bibinfo {author} {\bibfnamefont {L.}~\bibnamefont
  {Radzihovsky}}, \ and\ \bibinfo {author} {\bibfnamefont {V.}~\bibnamefont
  {Gurarie}},\ }\href@noop {} {\bibfield  {journal} {\bibinfo  {journal}
  {arxiv:1402.3737}\ } (\bibinfo {year} {2014})}\BibitemShut {NoStop}%
\bibitem [{\citenamefont {Biswas}\ and\ \citenamefont
  {Ryu}(2014)}]{Biswas2014}%
  \BibitemOpen
  \bibfield  {author} {\bibinfo {author} {\bibfnamefont {R.~R.}\ \bibnamefont
  {Biswas}}\ and\ \bibinfo {author} {\bibfnamefont {S.}~\bibnamefont {Ryu}},\
  }\href {\doibase 10.1103/PhysRevB.89.014205} {\bibfield  {journal} {\bibinfo
  {journal} {Phys. Rev. B}\ }\textbf {\bibinfo {volume} {89}},\ \bibinfo
  {pages} {014205} (\bibinfo {year} {2014})}\BibitemShut {NoStop}%
\bibitem [{\citenamefont {Ominato}\ and\ \citenamefont
  {Koshino}(2014)}]{Ominato2013}%
  \BibitemOpen
  \bibfield  {author} {\bibinfo {author} {\bibfnamefont {Y.}~\bibnamefont
  {Ominato}}\ and\ \bibinfo {author} {\bibfnamefont {M.}~\bibnamefont
  {Koshino}},\ }\href {\doibase 10.1103/PhysRevB.89.054202} {\bibfield
  {journal} {\bibinfo  {journal} {Phys. Rev. B}\ }\textbf {\bibinfo {volume}
  {89}},\ \bibinfo {pages} {054202} (\bibinfo {year} {2014})}\BibitemShut
  {NoStop}%
\bibitem [{\citenamefont {Nandkishore}\ \emph {et~al.}()\citenamefont
  {Nandkishore}, \citenamefont {Huse},\ and\ \citenamefont
  {Sondhi}}]{Nandkishore2013}%
  \BibitemOpen
  \bibfield  {author} {\bibinfo {author} {\bibfnamefont {R.}~\bibnamefont
  {Nandkishore}}, \bibinfo {author} {\bibfnamefont {D.}~\bibnamefont {Huse}}, \
  and\ \bibinfo {author} {\bibfnamefont {S.}~\bibnamefont {Sondhi}},\
  }\href@noop {} {\ }\Eprint {http://arxiv.org/abs/arXiv:1307.3252v2}
  {arXiv:1307.3252v2} \BibitemShut {NoStop}%
\bibitem [{\citenamefont {Hosur}\ \emph {et~al.}(2012)\citenamefont {Hosur},
  \citenamefont {Parameswaran},\ and\ \citenamefont {Vishwanath}}]{Hosur2012}%
  \BibitemOpen
  \bibfield  {author} {\bibinfo {author} {\bibfnamefont {P.}~\bibnamefont
  {Hosur}}, \bibinfo {author} {\bibfnamefont {S.~A.}\ \bibnamefont
  {Parameswaran}}, \ and\ \bibinfo {author} {\bibfnamefont {A.}~\bibnamefont
  {Vishwanath}},\ }\href {\doibase 10.1103/PhysRevLett.108.046602} {\bibfield
  {journal} {\bibinfo  {journal} {Phys. Rev. Lett.}\ }\textbf {\bibinfo
  {volume} {108}},\ \bibinfo {pages} {046602} (\bibinfo {year}
  {2012})}\BibitemShut {NoStop}%
\bibitem [{\citenamefont {Kobayashi}\ \emph {et~al.}(2014)\citenamefont
  {Kobayashi}, \citenamefont {Ohtsuki}, \citenamefont {Imura},\ and\
  \citenamefont {Herbut}}]{Kobayashi2014}%
  \BibitemOpen
  \bibfield  {author} {\bibinfo {author} {\bibfnamefont {K.}~\bibnamefont
  {Kobayashi}}, \bibinfo {author} {\bibfnamefont {T.}~\bibnamefont {Ohtsuki}},
  \bibinfo {author} {\bibfnamefont {K.-I.}\ \bibnamefont {Imura}}, \ and\
  \bibinfo {author} {\bibfnamefont {I.~F.}\ \bibnamefont {Herbut}},\ }\href
  {\doibase 10.1103/PhysRevLett.112.016402} {\bibfield  {journal} {\bibinfo
  {journal} {Phys. Rev. Lett.}\ }\textbf {\bibinfo {volume} {112}},\ \bibinfo
  {pages} {016402} (\bibinfo {year} {2014})}\BibitemShut {NoStop}%
\bibitem [{\citenamefont {Nielsen}\ and\ \citenamefont
  {Ninomiya}(1981)}]{Nielsen1981}%
  \BibitemOpen
  \bibfield  {author} {\bibinfo {author} {\bibfnamefont {H.}~\bibnamefont
  {Nielsen}}\ and\ \bibinfo {author} {\bibfnamefont {M.}~\bibnamefont
  {Ninomiya}},\ }\href@noop {} {\bibfield  {journal} {\bibinfo  {journal}
  {Nucl. Phys. B}\ }\textbf {\bibinfo {volume} {185}},\ \bibinfo {pages} {20}
  (\bibinfo {year} {1981})}\BibitemShut {NoStop}%
\bibitem [{\citenamefont {Bardarson}\ \emph {et~al.}(2007)\citenamefont
  {Bardarson}, \citenamefont {Tworzydlo}, \citenamefont {Brouwer},\ and\
  \citenamefont {Beenakker}}]{Bardarson2007}%
  \BibitemOpen
  \bibfield  {author} {\bibinfo {author} {\bibfnamefont {J.~H.}\ \bibnamefont
  {Bardarson}}, \bibinfo {author} {\bibfnamefont {J.}~\bibnamefont
  {Tworzydlo}}, \bibinfo {author} {\bibfnamefont {P.~W.}\ \bibnamefont
  {Brouwer}}, \ and\ \bibinfo {author} {\bibfnamefont {C.~W.~J.}\ \bibnamefont
  {Beenakker}},\ }\href {\doibase 10.1103/PhysRevLett.99.106801} {\bibfield
  {journal} {\bibinfo  {journal} {Phys. Rev. Lett.}\ }\textbf {\bibinfo
  {volume} {99}},\ \bibinfo {pages} {106801} (\bibinfo {year}
  {2007})}\BibitemShut {NoStop}%
\bibitem [{\citenamefont {Adam}\ \emph {et~al.}(2009)\citenamefont {Adam},
  \citenamefont {Brouwer},\ and\ \citenamefont {\mbox{Das Sarma}}}]{Adam2009}%
  \BibitemOpen
  \bibfield  {author} {\bibinfo {author} {\bibfnamefont {S.}~\bibnamefont
  {Adam}}, \bibinfo {author} {\bibfnamefont {P.~W.}\ \bibnamefont {Brouwer}}, \
  and\ \bibinfo {author} {\bibfnamefont {S.}~\bibnamefont {\mbox{Das Sarma}}},\
  }\href {\doibase 10.1103/PhysRevB.79.201404} {\bibfield  {journal} {\bibinfo
  {journal} {Phys. Rev. B}\ }\textbf {\bibinfo {volume} {79}},\ \bibinfo {eid}
  {201404} (\bibinfo {year} {2009})}\BibitemShut {NoStop}%
\bibitem [{\citenamefont {Tworzydlo}\ \emph {et~al.}(2006)\citenamefont
  {Tworzydlo}, \citenamefont {Trauzettel}, \citenamefont {Titov}, \citenamefont
  {Rycerz},\ and\ \citenamefont {Beenakker}}]{Tworzydlo2006}%
  \BibitemOpen
  \bibfield  {author} {\bibinfo {author} {\bibfnamefont {J.}~\bibnamefont
  {Tworzydlo}}, \bibinfo {author} {\bibfnamefont {B.}~\bibnamefont
  {Trauzettel}}, \bibinfo {author} {\bibfnamefont {M.}~\bibnamefont {Titov}},
  \bibinfo {author} {\bibfnamefont {A.}~\bibnamefont {Rycerz}}, \ and\ \bibinfo
  {author} {\bibfnamefont {C.~W.~J.}\ \bibnamefont {Beenakker}},\ }\href
  {\doibase 10.1103/PhysRevLett.96.246802} {\bibfield  {journal} {\bibinfo
  {journal} {Phys. Rev. Lett.}\ }\textbf {\bibinfo {volume} {96}},\ \bibinfo
  {pages} {246802} (\bibinfo {year} {2006})}\BibitemShut {NoStop}%
\bibitem [{\citenamefont {Goswami}\ and\ \citenamefont
  {Chakravarty}(2011)}]{Goswami2011}%
  \BibitemOpen
  \bibfield  {author} {\bibinfo {author} {\bibfnamefont {P.}~\bibnamefont
  {Goswami}}\ and\ \bibinfo {author} {\bibfnamefont {S.}~\bibnamefont
  {Chakravarty}},\ }\href {\doibase 10.1103/PhysRevLett.107.196803} {\bibfield
  {journal} {\bibinfo  {journal} {Phys. Rev. Lett.}\ }\textbf {\bibinfo
  {volume} {107}},\ \bibinfo {pages} {196803} (\bibinfo {year}
  {2011})}\BibitemShut {NoStop}%
\bibitem [{\citenamefont {Slevin}\ and\ \citenamefont
  {Ohtsuki}(1997)}]{Slevin1997}%
  \BibitemOpen
  \bibfield  {author} {\bibinfo {author} {\bibfnamefont {K.}~\bibnamefont
  {Slevin}}\ and\ \bibinfo {author} {\bibfnamefont {T.}~\bibnamefont
  {Ohtsuki}},\ }\href {\doibase 10.1103/PhysRevLett.78.4083} {\bibfield
  {journal} {\bibinfo  {journal} {Phys. Rev. Lett.}\ }\textbf {\bibinfo
  {volume} {78}},\ \bibinfo {pages} {4083} (\bibinfo {year}
  {1997})}\BibitemShut {NoStop}%
\bibitem [{Note1()}]{Note1}%
  \BibitemOpen
  \bibinfo {note} {The theory of Ref.\ \protect \rev@citealp {Ominato2013} can
  be applied to our calculations by making the substitutions
  $d_{0}\leftrightarrow \xi $ and $W\leftrightarrow K/2\pi $.}\BibitemShut
  {Stop}%
\bibitem [{\citenamefont {Rammer}(1998)}]{Rammer}%
  \BibitemOpen
  \bibfield  {author} {\bibinfo {author} {\bibfnamefont {J.}~\bibnamefont
  {Rammer}},\ }\href@noop {} {\emph {\bibinfo {title} {{Quantum Transport
  Theory}}}}\ (\bibinfo  {publisher} {Perseus Books},\ \bibinfo {year}
  {1998})\BibitemShut {NoStop}%
\bibitem [{Note2()}]{Note2}%
  \BibitemOpen
  \bibinfo {note} {Details of this calculation are given in the
  appendix.}\BibitemShut {Stop}%
\bibitem [{\citenamefont {Das~Sarma}\ \emph {et~al.}(2012)\citenamefont
  {Das~Sarma}, \citenamefont {Hwang},\ and\ \citenamefont {Li}}]{DasSarma2012}%
  \BibitemOpen
  \bibfield  {author} {\bibinfo {author} {\bibfnamefont {S.}~\bibnamefont
  {Das~Sarma}}, \bibinfo {author} {\bibfnamefont {E.~H.}\ \bibnamefont
  {Hwang}}, \ and\ \bibinfo {author} {\bibfnamefont {Q.}~\bibnamefont {Li}},\
  }\href {\doibase 10.1103/PhysRevB.85.195451} {\bibfield  {journal} {\bibinfo
  {journal} {Phys. Rev. B}\ }\textbf {\bibinfo {volume} {85}},\ \bibinfo
  {pages} {195451} (\bibinfo {year} {2012})}\BibitemShut {NoStop}%
\bibitem [{\citenamefont {Imry}(2002)}]{Imry2002}%
  \BibitemOpen
  \bibfield  {author} {\bibinfo {author} {\bibfnamefont {Y.}~\bibnamefont
  {Imry}},\ }\href@noop {} {\emph {\bibinfo {title} {Introduction to mesoscopic
  physics}}}\ (\bibinfo  {publisher} {Oxford University Press},\ \bibinfo
  {year} {2002})\BibitemShut {NoStop}%
\end{thebibliography}%

\newpage{}

\onecolumngrid

\section*{Appendix: Leading correction to SCBA self energy}

We compute the leading correction $\delta \Sigma$ to the SCBA self energy $\Sigma^{\rm SCBA}$ at zero momentum $\vk=0$ and zero energy $\omega=0$. The diagrammatic representation for the correction $\delta \Sigma(\vk,\omega)$ is shown in Fig.\ \ref{fig:diagrams}(b),
\begin{equation}
  \delta \Sigma(\vk,\omega) =
  \sum_{\vk_1,\vk_2} {\cal G}(\vk+\vk_1,\omega) {\cal G}(\vk+\vk_1+\vk_2,\omega)
  {\cal G}(\vk+\vk_2,\omega)
  \langle |U_{\vk_1}|^2 \rangle \langle |U_{\vk_2}|^2 \rangle,
\end{equation}
where ${\cal G}(\vk,\omega) = [\omega - H_0 - \Sigma(\vk,\omega)^{\rm SCBA}]^{-1}$ is the SCBA propagator and $U$ the disorder potential.
Taking the disorder correlator $\langle |U_{\vk}|^2 \rangle$ from Eq.\ (\ref{eq:Usimpl}), setting $\vk=0$, $\omega=0$, and replacing the summation over $\vk_1$ and $\vk_2$ by an integration one finds
\begin{eqnarray*}
  \delta \Sigma(0,0) & = & \KK'^{2}\xi^{2}\left(\hbar v\right)^{4}\int_{k_{1}<2/\xi}\frac{d\mathbf{k}_{1}}{(2\pi)^{3}}\int_{k_{2}< 2/\xi}\frac{d\mathbf{k}_{2}}{(2\pi)^{3}}{\cal G}(\mathbf{k}_{2},0)
  {\cal G}(\mathbf{k}_{1}+\mathbf{k}_{2},0) {\cal G}(\mathbf{k}_{1},0).
\end{eqnarray*}
Employing the identity $(a - \vb \cdot \vsigma)^{-1} = (a + \vb \cdot \vsigma)/(a^2-|\vb|^2)$ and substituting \cite{Ominato2013}
\begin{equation}
 \Sigma(0,0)^{\rm SCBA} = \frac{4 \pi i \hbar v} {\tilde \KK'  \xi},\ \
   \tilde \KK' = \frac{1}{1/\KK' - 1/\KKc'},
\end{equation}
for disorder strength $\KK' > \KKc'$, one finds that (for a positive helicity Weyl node) the single-particle propagator ${\cal G}$ is given by the expression
\begin{equation}
  {\cal G}(\vk,0) =
  \left( \frac{\xi}{2 \hbar v} \right)
  \frac{(2 \pi/\tilde \KK')i - (\xi/2) \vk \cdot \vsigma}{
  (2 \pi/\tilde \KK')^2 + (\xi/2)^2 k^2}.
\end{equation}
Switching to the dimensionless variables $\vx_{1,2} = \vk_{1,2} \xi/2$ we arrive at
\begin{eqnarray}
  \frac{\delta \Sigma(0,0)}{\Sigma(0,0)^{\rm SCBA}}
  &=& \frac{\KK'^2 \tilde \KK'}{32 \pi^7 i}
  \nonumber \\ && \mbox{} \times
  \int_{x_1 < 1} d\vx_1 \int_{x_2 < 1} d\vx_2
  \left( \frac{2 \pi i/\tilde \KK' - \vx_2 \cdot \vsigma}
    {(2 \pi/\tilde \KK')^2 + x_2^2} \right)
  \left( \frac{2 \pi i/\tilde \KK' - (\vx_1 + \vx_2) \cdot \vsigma}
    {(2 \pi/\tilde \KK')^2 + |\vx_1 + \vx_2|^2} \right)
  \left( \frac{2 \pi i/\tilde \KK' - \vx_1 \cdot \vsigma}
    {(2 \pi/\tilde \KK')^2 + x_1^2} \right).
\end{eqnarray}
Finally, after introducing polar coordinates for the integrations over $\vx_1$ and $\vx_2$ one finds after some standard manipulations
\begin{eqnarray}
  \frac{\delta \Sigma(0,0)}{\Sigma(0,0)^{\rm SCBA}} &=& 
  \frac{\KK'^{2}}{4\pi^{4}}
  \int_{0}^{1}dx_{1}\int_{0}^{1}dx_{2}
    \frac{x_1^2 x_2^2}{[(2 \pi/\tilde \KK')^2+x_{1}^{2}]
    [(2 \pi/\tilde \KK')^2+x_{2}^{2}]}
  \left\{ 6 - \left[ \frac{5}{2} (2 \pi/\tilde \KK')^2 + 
    \frac{1}{2} (x_1^2 + x_2^2) \right]
  \vphantom{\frac{M}{\sqrt{M_M^M}}} \right.
  \nonumber \\ && \mbox{} \left. \times
  \int_{0}^{\pi}d\zeta\int_{\zeta}^{\pi}d\theta
    \frac{\cos\zeta-\cos\theta}
    {\sqrt{[(2\pi/\tilde \KK')^2 + x_{1}^{2}+x_{2}^{2}+2x_{1}x_{2}\cos\zeta]
     [(2\pi/\tilde \KK')^2 + x_{1}^{2}+x_{2}^{2}+2x_{1}x_{2}\cos\theta]}}
  \right\}.
\end{eqnarray}
Numerical evaluation of the fourfold integral for $\KK'$ in the vicinity
of the critical disorder strength $\KKc'$ then results in the estimate
(\ref{eq:estimate}) quoted in the main text.

\end{document}